\documentclass[12pt,preprint]{aastex}
\def\lax {\ifmmode{_<\atop^{\sim}}\else{${_<\atop^{\sim}}$}\fi}
\def\gax {\ifmmode{_>\atop^{\sim}}\else{${_>\atop^{\sim}}$}\fi}

\begin{document}

\title{Outflows Near An Accreting Black Hole: Ionization and
Temperature Structures}
% Why Andy Fabian Should Return the Rossi Prize!}

\author{J. Martin Laming\altaffilmark{1} and Lev Titarchuk\altaffilmark{2}}

\altaffiltext{1}{US Naval Research
Laboratory, Code 7674L, Washington, DC 20375-5352; jlaming@ssd5.nrl.navy.mil }
%\altaffiltext{2}{F\'ed\'eration APC, Coll\`ege de France, 75231 Paris, France}
\altaffiltext{2}{George Mason University/Center for Earth
Observing and Space Research, Fairfax, VA 22030; and US Naval Research
Laboratory, Code 7655, Washington, DC 20375-5352; ltitarchuk@ssd5.nrl.navy.mil }
%\altaffiltext{4}{NASA/ Goddard Space Flight Center, code 660, Greenbelt
%MD 20771, lev@lheapop.gsfc.nasa.gov}

\begin{abstract}
We calculate the temperature and ionization balance in an outflow from an accreting
black hole under illumination by hard radiation from the central object.
Electron scattering of the Fe $K_{\alpha}$ photons within the highly ionized
expanding flow leads to a decrease of their energy (redshift) which is of
first order in $v/c$, when $v$ is the outflow velocity ($v$ is much less
than  the speed of light $c$). {\it This photon redshift is  an intrinsic
property of  any outflow for which divergence is positive}.
We also find  that the equivalent widths of red skewed Fe K$\alpha$ originated
in the wind is order of keV.
We  conclude that  redshifted lines are intrinsic properties of the powerful
outflows that are observed  in many compact objects.
Downscattering of the primary line photons generated in the outflow
(a more natural and probable mechanism than  the general relativistic effects
in the innermost part of the accretion flow) leads to the formation of
red-skewed lines.
\end{abstract}

\section{Introduction}
There is a great need for a coherent physical picture predicting
how X-ray spectra are produced
near galactic and extragalactic black hole (BH) sources.
Elvis (2003) points out that the BH-disk-jet theory is presently  a ``high''
theory, dealing only with overall energetics, and describes a naked object devoid
of any of the
veiling gas that presumably creates the observable spectral continuum and atomic
features.
This ``high'' theory  fails to explain the various  correlations between timing and
spectral features observed in BH sources  [for example, see details of these
correlations in  Mauche (2002); Titarchuk \& Wood (2002);
Titarchuk \& Fiorito (2004)].
Elvis concludes that the community needs a ``low'' theory with sufficient
detail to predict the
emission (and presumably timing) and absorption phenomenology of BHs.

%Vaughan \& Fabian (2003) argue that X-ray spectroscopy holds great promise for probing physics close
%to accreting BHs. They hypothesize that the effects of strong gravity, such as gravitational redshift and
%light bending near BH are responsible for and  should imprint characteristic signatures on the emerging X-ray spectrum
%like the  iron line feaure.
Many authors [e.g. Tanaka et al. (1995); Nandra et al. (1997); Wilms et al. (2001);
Uttley et al. (2003); Miller et al. (2004)]
have found unusual curvature (red-skewed  features) at energies between 2 and 8 keV in X-ray spectra of a number of
galactic and extragalactic BH sources. Significantly, the fluxes of these features remain nearly {\it constant}
despite the large changes in the continuum flux from the central object
as shown in the observations of MCG-6-30-15, NGC 4051 and others [Markovitz,
Edelson, \& Vaughan (2003), hereafter MEV03], suggesting an origin in the outer regions of the
accretion flow.  In fact, MEV03 discuss a possibility of explanation of 
 the lack of correlation between the continuum and the line using the ionized disk skin model by Nayakshin
 (2000) and Nayakshin \& Kallman (2001) (see also Ross, Fabian \& Young 1999). This model is consistent with 
 the prediction that both broad and narrow K$_{\alpha}$ lines should track continuum variations on time
 scales of months to years. However such correlated behavior is not seen through the entire sample. 

%Photons can lose energy in scattering with cold electrons due to recoil (see e.g. Sunyaev \& Titarchuk 1980,hereafter
%ST80). In Figure 10 in ST80 the resulting spectra of the passage of the
%photons through a cold  medium for a power-law continuum are presented.
%One can clear see a  broad bump in the range of tens of keV  for the emerging spectrum when the optical depth
%of the cloud is about 5 and higher.
In this paper we further discuss a model for redshifting of emission lines
by repeated electron scattering in a diverging outflow or wind. The
basic mechanism is described by Laurent \& Titarchuk (2004). It is
distinct from the proposals for redshifts previously associated with
outflows, Comptonization (in a low temperature electron cloud, see
Sunyaev \& Titarchuk 1980, for details) and the general accelerating outflow
model of
Sobolev (1957), both of which were correctly ruled out by Fabian et al.
(1995) in connection with the redshifted Fe K$\alpha$ in MCG-6-30-15.
Comptonization would produce an unobserved spectral break
at energy $E\sim m_ec^2/\tau_{\rm T}^2\sim 20$ keV where $\tau_{\rm T}$ is the outflow 
Thomson optical depth,
The accelerating outflow model required
optically thick clouds moving with velocities $v\sim 100~000$ km s$^{-1}$.
They argue that the major problem with this outflow model is that MCG-6-30-15
has a warm
absorber. The OVII edge in the absorber indicates that the flow along line of
sight is
less than about 5000 km s$^{-1}$ (Fabian et al. 1994), which seriously
conflicts with outflow of about 100 000 km s$^{-1}$.
Laurent \& Titarchuk (2004)  confirm
that velocity of the warm absorber outflow in MCG-6-30-1 is about 5000 km s$^{-1}$
using Monte Carlo simulations of the $K_{\alpha}$ line
propagation in   the constant-velocity wind.
They present the results of  fitting the wind model to the XMM (EPIC) data of
Wilms et al. (2001) obtained during the observation of
 MCG-6-30-15. These observations were focused on the broad Fe K$\alpha$ line at
$\sim 6.4$ keV. The continuum has been fitted with a power law of
index 1.8 and  the residuals have been fitted using the wind  model.
The best-fit model parameters they found are $E_{ph} = 6.5$ keV, ~$kT = 0.1$ keV,
$\tau_{\rm T} =1.7, ~\beta=v/c= 0.02$ where $E_{ph}=h\nu_{ph}$ is energy of the
primary $K_\alpha$ photons, and  $kT$    is the temperature of  the wind respectively.
Downscattering of the primary line photons in the outflow  leads to the formation of
red-skewed lines.  This effect is seen even when the  outflow optical depth
$\tau_{\rm T}$ is about 1-2 and $\beta\gax v/c\sim 0.02$. 

Turolla et al. (1996) were the first to analyze the radiative transfer in
expanding (diverging) flow using a Fokker-Planck equation derived by Blandford and
Payne (1981). A thorough analysis  and review of the diffusion theory of photon propagation
in an optically thick fluid in a bulk outflow has been provided by Titarchuk, Kazanas \& Becker (2003), hereafter TKB03.
They show that  the iron line is produced in an effectively optically thick medium.
Its red wing is the result of multiple scattering, each scattering producing a first
order $v/c-$redshift (see Fig.1 for a simple explanation of the outflow  redshift effect).
This process produces a red wing to the line without any particularly fine-tuned geometric arrangement.
The TKB03 results are obtained using the Fokker-Planck equation [Blandford \& Payne (1981)]
which   is valid for $\beta=v/c\ll 1$ and $\tau_{eff}>1$. It is worth noting that recently, 
using this Fokker-Planck technique, Titarchuk \& Shrader (2004) presented a detailed study of  
the outflow downscattering effect in the continuum.
  The Monte Carlo (MC)
method does not have these limitations in terms of $\beta$ and $\tau_{eff}$.
Using a Monte Carlo method Laurent \& Titarchuk (2004), hereafter LT04,
have made extensive radiative transfer simulations
%of  the propagation of monochromatic photons
in a  bulk  outflow from a compact object (black hole or neutron
star).
% They apply  generic assumptions regarding the distribution of these
%photons within the outflow,  the plasma temperature $kT$, and the optical depth  $\tau_{\rm T}$.
They find that electron scattering of the photon within the expanding flow leads to a decrease of its energy which is of
first order in $v/c$.  LT04 demonstrate that the emergent line iron profile is closely
related to the time distribution of photons diffusing through the flow
(the light curve) and exhibits a broad redshifted feature.
LT04 find so called  negative time lags, related to
the time dependence of the photon energy losses from propagating out of the flow.
%We fitted our model line profiles to the observations
%using  four free parameters, $\beta=v/c$, $\tau$, $kT_e$ and the original line photon energy $E_0$.
%In fact, the fits  weakly depend on the outflow plasma temperature for  $kT_e<1$ keV.
%Also to fit the data we assume that the source of the line photons is located at
% the bottom of the outflow.
%As a result of our simulations
Thus one can  conclude that  redshifted lines are intrinsic properties of the powerful
outflows that are observed  in many compact objects.
%Before proceeding we must also note the concerns of some authors, e.g. Miller et al. (2004),
%that the hot disk inner-disk
%component and high frequency QPOs ($\sim$few times
%100 Hz) seen in accreting Black Hole (BH) at the highest inferred accretion rates
%would not be visible through an ``optically thick'' outflow.
%Those authors thus rule out any possibility of ambient spectral reprocessing and
%distortion of the iron line feature, such as broadening and red-wing enhancement,
%by the outflow (see Titarchuk, Kazanas \& Becker 2003, hereafter TKB).  On the other hand, we
%infer the outflow Thomson optical depth $\tau\sim 1-2$  from parameter fitting using  XMM  and ASCA measurements
%of red-shifted iron lines (see \S 3).
%Thus the observer probably sees the radiation of the BH central source through
%the ``haze" of the moderate optical depth. Furthermore Titarchuk, Cui \& Wood (2002), hereafter TCW02
%give a precise model for the loss of the modulation due photon scattering.
%It follows from TCW02 that  $\tau$
%would need to be around 16 and higher in order to  suppress  QPO amplitude
%of frequency 100 Hz [see formula  (5) in TCW02]. But this very optically thick outflow is  ruled out by observations.

In this paper we argue that the Fe $K_{\alpha}$ line is naturally
produced in an external outflow illuminated by the hard X-ray radiation
coming from the central source (BH), and the red-skewed features are
the result of scattering of iron line photons  in the outflow.
We offer a model in which the hard radiation of the central object illuminates and heats the outflow region
originated in the outskirts of the disk (well outside the innermost part of the accretion disk near the BH).
% The line photons are generated and scattered in the outflow.
%The outflow is heated mainly by the hard radiation of the
%central object.
We present details of self-consistent calculations of atomic features
and temperature structure within the ouflow. The result of these calculations for $\beta=0$
 gives us the temperature and ionization structure within the disk.

%An increase of photon source concentration towards the outflow bottom
% (i.e., producing a larger fraction of the photons  deep into the flow)
% leads to much broader profiles which can  exhibit
%a shoulder or a secondary minimum [not unlike the Fe K$\alpha$ line feature
%that was interpreted as evidence for infalling matter in NGC 3516 (Nandra et
%al. 1999)]. The origin of this feature is well understood within the TKB model:
%the narrow line  corresponds to photons which escape  without  scattering,
%while the broader, redshifted profile is formed by the photons
%trapped in the flow which diffuse their way out through
%the expanding flow.
%It is also pointed out that photon scattering in diverging flows
%produces  negative time lags in the spectra (see Figs. 2-3 and formula 21 in TKB).

In \S 2 we present a description of the outflow  illumination model
and provide the main idea of the photon frequency shift due
to electron scattering  in the diverging flow (outflow) and converging
flow (inflow).
We show the results of calculations  of temperature and ionization  balance in the outflow and evaluation
of the equivalent widths of Fe K$\alpha$  as a function of the ionization parameter
for various incident X-ray spectral distributions and  outflow bulk velocity  $\beta$, in \S 3.
Discussion and conclusions follow in \S 4.

\section{Outflow Illumination by Black Hole X-ray Radiation: Effect of the $K_{\alpha}-$ Photon Outflow Redshift and
 Suppression of the Blue Wing of $K_{\alpha}-$line}
%\subsection{}
Our basic scenario is illustrated in Figure 1. The
wind originates at a distance $R_{inner}$ from the central black hole,
and is of a density such as to give a Thomson scattering optical depth $\tau_{\rm T}$
close to unity far from the black hole. The optical depth in the Fe K
continuum is about 1-3 times higher than that due to electron scattering
(assuming a solar abundance of Fe, and depending on charge state, see
Kallman et al. 1994),
and so Fe K$\alpha$ formed by inner shell ionization of Fe ions in the
outflow by the continuum from the central black hole only comes from a
smaller inner region. So long as $R\left(\tau _{\rm Fe K}\sim 1\right) <
R\left(\tau _{\rm T}\sim 1\right)$, the Fe K$\alpha$ equivalent
width is insensitive to the Fe abundance assumed. A lower Fe abundance
increases $R\left(\tau _{\rm Fe K}\sim 1\right)$, and consequently
Fe K$\alpha$ forms over a larger volume. This figure also illustrates the
redshift of photons in the diverging flow. A photon emitted near the inner boundary and subsequently
scattered by an electron  moving with velocity ${\bf v}_1$, is
received by an electron moving with velocity ${\bf v}_2$ as shown with
frequency $\nu_2 = \nu_1[1+({\bf v}_1-{\bf v}_2)
\cdot{\bf n}/c]$ where ${\bf n}$ is a unit vector along the
path of the photon at the  scattering point. In a diverging flow
$\left({\bf v}_1-{\bf v}_2\right)\cdot{\bf n}/c <0$ and photons are
successively redshifted, until scattered to an observer at infinity.
In a converging flow
$({\bf v}_1-{\bf v}_2)\cdot{\bf n}/c >0$ and photons are
blueshifted. 
%Small blueshifts may also occur in the diverging flow due to
%the kinetic nature of the electrons, as in the calculation of
%LT04 (hot electrons may have thermal
%speeds in excess of the wind outflow velocity). Any Fe K photon blueshifted
%in this manner to an energy above the Fe ionization potential is likely
%to be absorbed, thus suppressing the blue wing of the Fe K line, for energies
%higher than 7.1- 9 keV, depending on Fe charge state.
%Recent results by Kallman et al. (2004) show that the iron photo-absorption opacity $\sigma_K$ is about few times
%the Thomson opacity, $\sigma_{\rm T}$.
%Namely,  any photon
%around the K-absorption edge energy must be absorbed.
%It is worth noting, that these edges are seen with the strong
%K$_\alpha$ lines presumably from outflows during  X-ray superbursts on neutron
%stars (see  Figs. 5 and 9 in Strohmayer \& Brown 2002).
Any photon with energy
about 7 keV and higher interacting with  outflow plasma is more likely
to be absorbed by the flow and be reemitted at energies about 6.4-6.6 keV
(depending on the ionization stage of the flow) instead of being
scattered by electrons there.
This photo-absorption effect
is particularly important
% this partiof the referee makes even
in the view of {\it the main claim of this work as a suppression of the blue wing at the expense of the $K_\alpha$
emission at energies about 6.4-6.6 keV}.
% and might even be visible in the data (see LT04).

\section{Temperature and Ionization Balance in the Outflow}
%\subsection{Formulation of the Problem}
%\vscape{0.25in}
%~~~~~~~~~~~~~~~~~~~~~

\centerline{\it Formulation of the Problem}

The temperature and ionization balance
 balance in the outflow is determined by
seeking the temperature at which the outflowing gas attains
photoionization-recombination equilibrium.
The gas is heated by Compton
scattering and photoionizations by photons from the central compact object,
and cooled by radiation, ionization and adiabatic expansion losses. We
take collisional ionization and recombination rates from \citet{mazzotta98},
photoionization cross sections from \citet{verner96} and radiative power losses from
\citet{summers79}. For the
Fe K photoionizations we modify the Verner et al. (1996)
cross sections by a multiplicative
factor in the range 1-3 depending on charge state to bring these cross
sections into better agreement with the more recent work of Kallman et al.
(2004).
It is evident that the temperature and ionization structure in the disk due to illumination of the hard radiation
can be obtained as a result of this solution for $\beta=v/c=0$.

%In fact, the solution of this  problem is  general for calculation of the  temperature and ionization balance in the target
%illuminated by the hard radiation and it can have a lot of applications beyond of Astrophysics
%(for example, in  Safety Radiation Physics).
%\vscape{0.25in}
~~~~~~~~~~~~~~~~~~~~~~~~~~~~

\centerline{\it Results of Calculations}

%\subsection{Results of Calculations}

For each inner wind radius, we define a density which gives an electron
scattering opacity in the wind  $\tau_{\rm T}$, thus density
$n\propto \tau_{\rm T}/R_{inner}$,
where $R_{inner}$ is the inner radius of the wind.
In outer portions of the wind, the density varies as $n\propto 1/R^2$,
so the ionization parameter, $L/nR^2$, remains approximately
constant. The Compton and photoelectric heating rates, approximately
proportional to $n/R^2 \propto 1/R^4$, fall off slightly faster
than the cooling rate (proportional to $n/R\propto 1/R^3$ for adiabatic
expansion, proportional to $n^2\propto 1/R^4$ for radiative losses)
in a realistic wind. The temperature we calculate
at the inner edge of the wind is the highest temperature in the flow in cases
where adiabatic expansion dominates the cooling. Where radiative cooling
is more important, the heating and cooling are more evenly balanced, and
the temperature we calculate is likely to be approximately correct
throughout the flow. Detailed radiation transfer modelling
accounting for successive absorption and scattering of the incident
hard spectrum to improve upon this estimate is beyond the scope of this
work.

In Figure 2  we plot the run of plasma temperature against
initial wind radius for a variety of models. The top panel shows
temperature (solid line, to be read on the left hand y-axis) and
average Fe charge state (dashed line, to read read on the right hand
y-axis) for
an outflow with $\beta = v/c = 0.1$ and for incident spectra
$I(\nu)\propto \nu ^{-\alpha}\exp(-h\nu /2kT)$ with
$\alpha=0.25$, 0.5, 0.75 and $kT=50$ keV. The inverse of the ionization parameter
$nR^2/L\simeq \tau_{\rm T}R_{inner}/L_{40}$ on this plot, where
$L_{40}$ is the source luminosity in $10^{40}$ ergs s$^{-1}$. For $\tau_{\rm T}R_{inner}/L_{40}<10^{12.8}$ cm,
Fe is in the He-like charge state or higher, and adiabatic
expansion dominates the cooling. As L-shell ions Fe XVII-XXIV start to form,
radiative cooling begins to dominate and the temperature drops more
steeply, with regions of thermal instability between Fe charge states
16+ and 13+. Harder spectra (i.e. lower $\alpha$) with more photons at
high energies have slightly lower photoionization rates for charge
states Fe$^{23+}$ and below (for the same luminosity), and so under these
conditions the outflowing plasma recombines and cools slightly faster.
The lower panel shows similar plots, but now with $\alpha=0.5$
and $\beta =v/c =0.02$, 0.05, 0.1, and 0.25. Faster outflows give more
adiabatic cooling and so cool and recombine at lower radii than do the
lower outflows.

Figure 3  shows the equivalent widths of Fe K$\alpha$ (dashed lines,
to be read on the right hand axis) for the
same sets of parameters as for Figure 2. The solid lines to be read on the
left hand axis give the total opacity in the outflow at 4 keV, in the
form $\exp\left(-opacity\right)$. At small
radii the equivalent width is essentially zero because Fe is in charge
state 24 or higher, with no L shell electrons available to fill a K
shell vacancy. As Fe recombines at larger radii, the equivalent width
increases. The predicted equivalent width at large inner radii of about 5 keV
is significantly larger than that observed of about 1 keV. However the
total wind opacity at these radii is such that most of the Fe K line
would be absorbed further out. Only for $R_{inner}\sim 10^{13}L_{40}$
and consequently for the wind temperatures about a few times $10^6$ K (see Figs. 2-3)
to where the equivalent width is approximately 1 keV is the opacity sufficiently
small to allow the line to be observed. We also note that this opacity is
likely to have the effect of suppressing any Fe L shell features between
1 and 2 keV photon energy.

\section{ Discussion and Conclusions}
We find the self-consistent temperature and ionization structure
of the wind shell as a function of the parameter (radius/luminosity,
$\tau_{\rm T} R_{inner}/L_{40}$, 
 where $R_{inner}$ is
the inner outflow radius and $L_{40}$ is X-ray luminosity in units of
$10^{40}$ erg s$^{-1}$) for a given shell Thomson optical depth $\tau_{\rm T}$.
%of order unity.
It is evident that the ionization parameter $L_{40}/nR^2$ is constant
through the shell (i.e.
for $R>R_{inner}$) if the velocity of the wind is
constant through the flow.
The constraints on the Fe ionization balance and the opacity in the wind
dictate a possible range of interest
of  $\tau_{\rm T}R_{inner}/L_{40}$ values around $10^{13}$ cm.
Thus our solution allows to determine the size of the shell base for a given
luminosity for which the red-skewed line is observed. For a typical
low/hard-state luminosities of a few times $10^{36-37}$ erg s$^{-1}$
when the red-skewed line is really observed in galactic black holes the
inner radius of the shell is  a few times $10^{9-10}$ cm for the optical depth of
the wind of order unity.
It means that the inner radius of the wind shell is about $10^{3-4}$ of
Schwarzschild radii.
One can obtain the similar size of the wind shell in Schwarzschild radii units for extragalactic sources because the luminosity and
the Schwarzschild radius is linearly scaled with BH mass.

It is important to note that if Thomson optical depth of outflow is of order of unity then the mass outflow rate $\dot M_{out}$
is of order of the Eddigton mass accretion rate $\dot M_{\rm Edd}$ and higher (see e.g. formula 4 in King \& Pounds 2003, hereafter KP03).
King (2003) and KP03
present strong arguments that powerful mass outflows from
Eddington-limited accreting compact objects appear to be a very widespread phenomenon. They further argue that
they may provide the soft excess observed in quasars and ULXs, and imply that such objects have a major effect on their surroundings.
Recent {\it XMM-Newton} observations of bright quasars (Pounds et al., 2003a,b; Reeves et al., 2003) give strong evidence
for powerful outflows from the nucleus with mass rates $\dot M_{out}\sim \dot M_{\odot}$ yr$^{-1}\sim \dot M_{\rm Edd}$ and
velocity $v\sim0.1c$ (i.e. $\beta=0.1$) in the form of blueshifted X-ray absorption lines. These outflows closely resemble those resently inferred in
a set of ultraluminous X-ray sources with extremely soft spectral components (Mukai et al. 2003; Fabbiano et al., 2003).
Furthermore, if the outflow optical depth is really about one in the low/hard state (when the X-ray luminosity is much less the Eddington luminosity)
one can conclude that in the low/hard state the disk mass accretion rate  is only very small fraction of the outflow mass rate.
In fact, this ratio of the  accretion and outflow rates  was predicted
by Blandford and Begelman (1999)  who developed the pure hydrodynamical model (so called ``ADIOS" model).
%For the outflow optical depth about one the ratio of outflow mass rate to the Eddigton mass accretion rate $\dot M_{Edd}$
%($\dot M_{Edd}=L_{Edd}/c^2$) is  about 7 independently of the mass of the central object.
%This estimate is made for  the hard state luminosity which is of order
%1\% of Eddington (this is really an observed quantity).
%Thus if the mass accretion rate in the hard state is only 10\% of Eddington then one can obtain that the ratio
%of outflow mass rate to mass accretion rate is about $70(\beta/0.1)$.
%In terms of the gravitational energy release in the system $E_g$ and its relation
%to the kinetic energy of the outflow $E_{outflow}$ in the hard state (when the red-skewed line is observed)
%one can conclude  that $E_{g}/E_{outflow}$  is about $100(\beta/0.1)^2$ independently of the mass of compact object.
%This issue is  particularly important for observational implications of  ADIOS
%(BB99) and ADAF (Narayan and Yi 1995) models widely discussed in the literature.
The main point of the ADIOS model is that  only a tiny fraction of the gas
supplied actually falls on to black hole.  This is precisely what  we obtain
using the diverging flow model.

We conclude that  the range  of the parameter $ \tau_{\rm T} R_{inner}/L_{40}$
(proportional to the inverse of the so called ``ionization parameter''
used in the literature) is
about $10^{13}$ when the observed $K_{\alpha}$ lines of the $\sim 1$ keV
equivalent widths are presumably produced in the
wind. The wind temperature (in such a case) is about a few times $10^6$ K. Thus our study shows that the strong iron line with its red-skewed
feature can be generated in
the relatively cold extended region far away from the source of the illuminating photons
(of order $10^{3-4}$ of Schwarzschild radii).

From the other hand one can argue that the ``standard'' cold disk plus hot corona
explanation for the broad lines is ``very robust (it needs only the
cold disk and a hot corona)'' and one does not need any other explanation for this effect.
 In fact, the so called ``cold disk'' can survive under the hot corona only if the corona is situated at least
$10^3-10^4/\tau_{\rm T}$ Schwarzschild radii where $\tau_{\rm T}$ is the Thomson optical depth of the disk.
If corona is much closer the disk as a target illuminated by the hard radiation of the corona is very hot
and finally evaporated. It is evident from  Figs. 2-3, that temperature  goes to $10^9$ K and  and
K$_{\alpha}$ equivalent width goes to zero  as $\tau_{\rm T}R_{inner}/L_{40}\ll10^{13}$ cm for $\beta< 0.05$.
%This result has nothing to do with any modeling it is just a result
%of the Basic Physics of interaction of the hard radiation with the target.  These type of calculations are around in Nuclear Physics
%last 60 years.
Other points regarding the ``robustness'' of the standard interpretation
are: even the line is produced in the disk  how one can see K$_\alpha$ through the corona
if the corona optical depth $\tau_{cor}$ is of order of one
and the corona electron temperature is around 60 keV?
The directed radiation of the K$_{\alpha}$ emission is attenuated exponentially as
$\exp(-\tau_{cor}/\mu)$ (at least) $< 1/3$
where $\mu$ is  the cosine of the inclination angle.
But the scattered component is completely smeared out - the relative energy change of the line photon at any
scattering is $<\Delta E/E>=4kT/m_ec^2=0.5$ (see e.g. Sunyaev \& Titarchuk 1980).
% This is a basic things of the Radiative Transfer.
But even if the disk line photons manage to come to the observer one would see them with very prominent
blue wing, which is a result of first order $v/c$ effect,
as the gravitational redshift is an effect of second order with respect to v/c, i.e  $R_{\rm S}/R \sim (v/c)^2$
where $R_{\rm S}$ is Schwarzschild radius.
%The combination of the red-skewed features along with the soft time lags
%that strongly depends on $\beta$, $\tau$ and $kT_e$ are intrinsic signatures of any diverging outflow.
The demonstrated application of our outflow model to  data   points out a potential
powerful spectral  diagnostic for probes of the outflow-central object connection in Galactic and
extragalactic BH sources.
%In the future, the {\it ASTRO-E2} and  {Constellation X} missions will make possible
%detailed studies  of these spectral features of iron lines along with their intrinsic timing properties.

J.M.L. is supported by
basic research funds of the Office of Naval Research.
L.T. acknowledges the support of this work by the Center for Earth Observing
and Space Research of the George Mason University.
%We also acknowledge the thorough analysis
%of this paper by the referee and his/her constructive and interesting suggestions.

\newpage
%\begin{figure}
%\epsscale{0.45}
%\plotone{f1.eps}
% fig 1
%\caption{
%}
%\label{fig1}
%\end{figure}

\begin{figure}
\includegraphics[width=6.in,height=4.2in,angle=0]{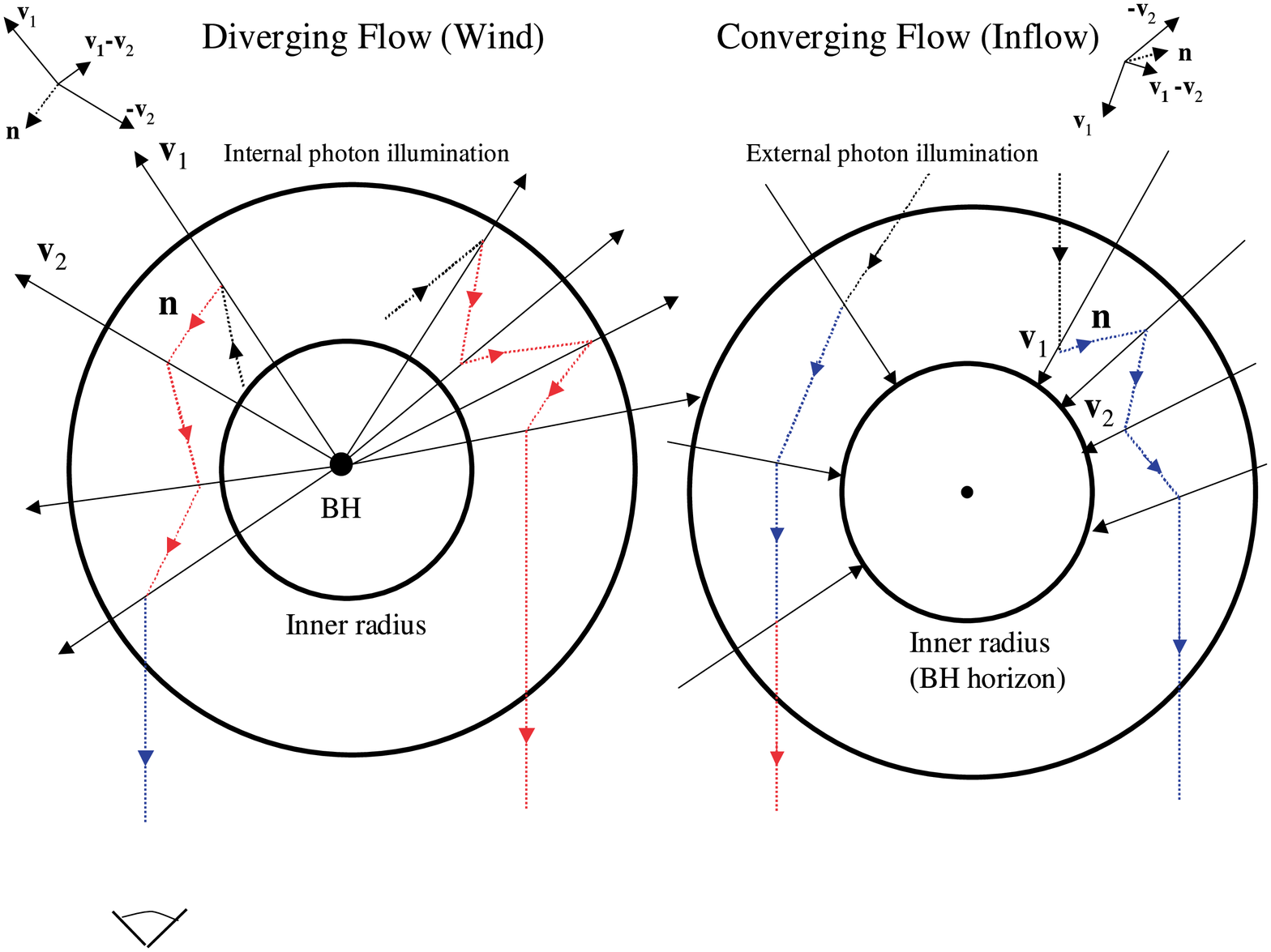}
\caption{On the left side: Schematic diagram showing wind geometry. The outflow (wind)
originates at the inner radius. The radii where the optical depth in
the Fe K continuum and the electron scattering optical depth are
of order unity. A photon emitted near the inner boundary and subsequently scattered
 by an electron moving with velocity ${\bf v}_1$, is
received by an electron moving with velocity ${\bf v}_2$ as shown.  The change in frequency is
 $\nu _2 = \nu _1\left(1+\left({\bf v}_1-{\bf v}_2\right)
\cdot{\bf n}/c\right)$ where ${\bf n}$ is a unit vector along the
path of the photon scattered at (b). In a diverging flow
$\left({\bf v}_1-{\bf v}_2\right)\cdot{\bf n}/c <0$ and photons are
successively redshifted, until scattered to an observer at infinity.
The color of photon path indicates the frequency shift in the rest frame of the receiver
(electron or the Earth observer).
On the right side: In a converging flow
$\left({\bf v}_1-{\bf v}_2\right)\cdot{\bf n}/c >0$ and photons are
blueshifted.
 }
\end{figure}
\newpage
\begin{figure}
\includegraphics[width=5.in,height=7.in,angle=0]{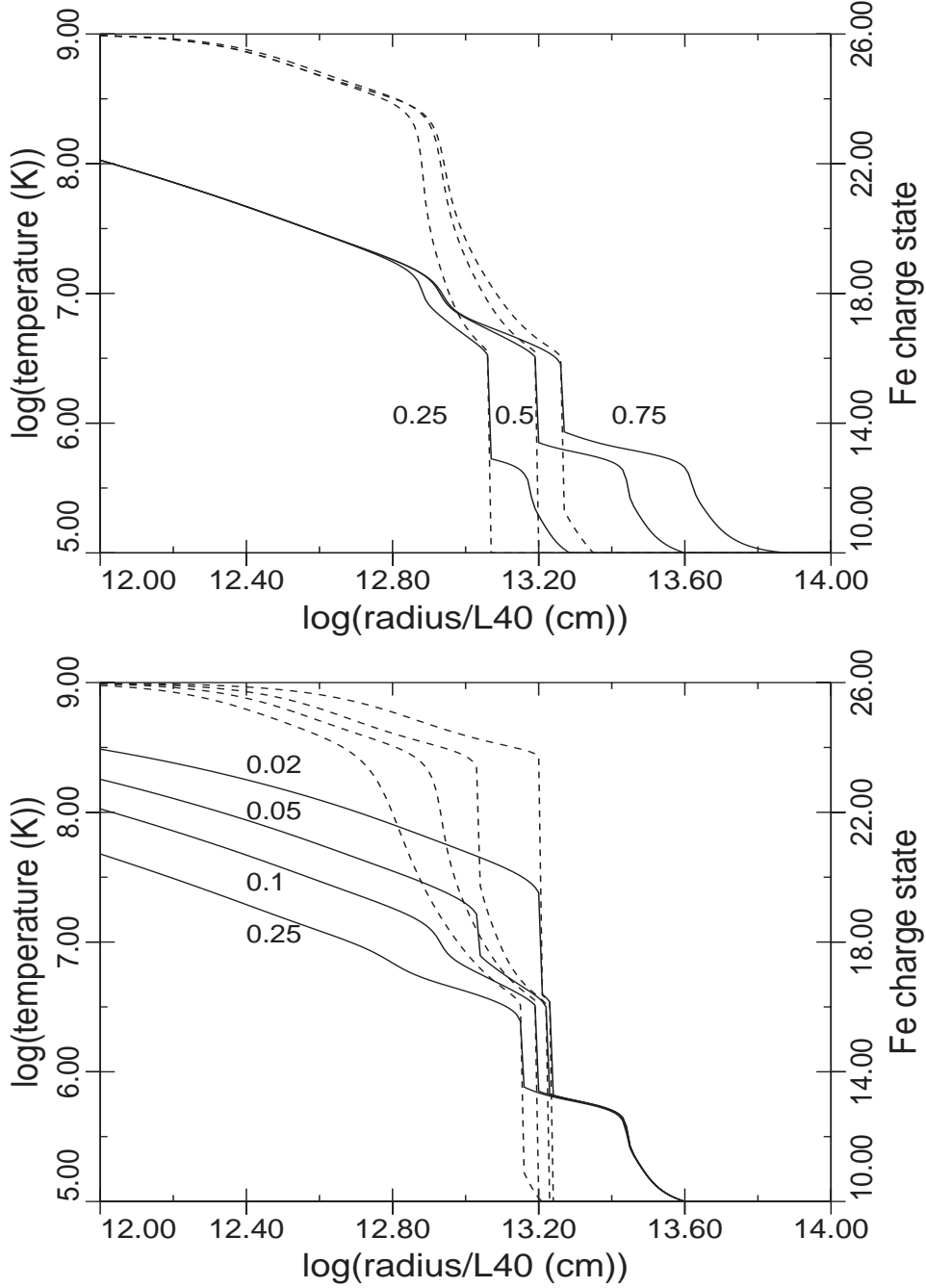}
\caption{Plot of temperature at $R_{inner}$ against $R_{inner}$ for an
accretion wind (solid lines).
The  inner radius $R_{inner}$ scales with the the central source luminosity and the optical depth of the wind shell
$\tau_{\rm T}$.
For any $\tau_{\rm T}\neq1$ the temperature is a function of $\tau_{\rm T} R_{inner}/L_{40}$.
The top panel shows results for expansion speed $\beta = v/c = 0.1$ and
spectral indices $\alpha=0.25$, 0.5, and 0.75. Bottom panel shows results for
$\alpha=0.5$ and $\beta = v/c =0.02$, 0.05, 0.1, and 0.25. The dashed lines
show the average Fe charge state (to be read on the right hand axis).}
\end{figure}
\newpage
\begin{figure}
\includegraphics[width=5.in,height=7in,angle=0]{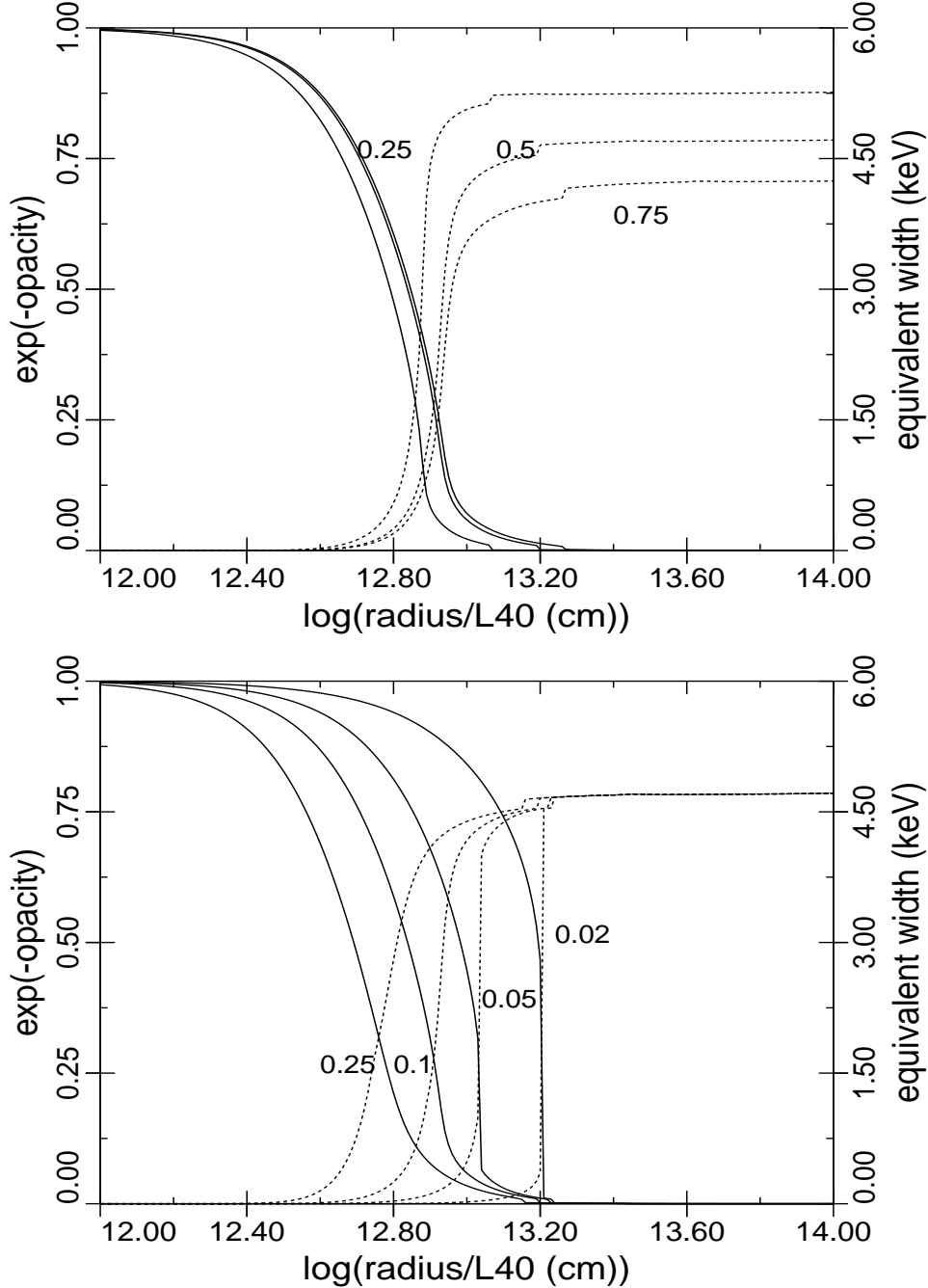}
\caption{Plots of the equivalent width of Fe K$\alpha$ (dash lines) produced by
inner shell ionization of Fe in an accretion wind by continuum from
the central source, for the same parameters as is Figure 2. At small
radii the equivalent width is essentially zero because Fe is in charge
state 24 or higher, with no L shell electrons available to fill a K
shell vacancy. As Fe recombines at larger radii, the equivalent width
increases. The solid lines to be read on the
left hand axis give the total opacity in the outflow at 4 keV, in the
form $\exp\left(-opacity\right)$. The predicted equivalent width at large radii of about 5 keV
is significantly larger than that observed of about 1 keV. However the
total wind opacity at these radii is such that most of the Fe K line
would be absorbed further out. Only for $\tau_{rm T} R_{inner}/L_{40}$ out to where the
equivalent width is approximately 1 keV is the opacity sufficiently
small to allow the line to be observed.
}
\end{figure}
%\newpage

\end{document}